\UseRawInputEncoding
\documentclass[aps,preprintnumbers,superscriptaddress,showpacs,UTF8]{revtex4}
\usepackage{epsfig}
\usepackage{psfrag}
\usepackage{amsfonts}
\usepackage{graphicx}
\usepackage{dcolumn}
\usepackage{bm}

\begin{document}

\title{Entropic destruction of heavy quarkonium in a rotating hot and dense medium from holography}

\author{Ping-ping Wu}
\email{wupp@hpu.edu.cn} \affiliation{School of Physics and
Electronic Information Engineering, Henan Polytechnic University,
Jiaozuo 454000, China}

\author{Zi-qiang Zhang}
\email{zhangzq@cug.edu.cn} \affiliation{School of Mathematics and
Physics, China University of Geosciences, Wuhan 430074, China}

\author{Xiangrong Zhu}
\email{xrongzhu@zjhu.edu.cn} \affiliation{School of Science,
Huzhou University, Huzhou 313000, China}

\begin{abstract}
Previous studies have indicated that the peak of the quarkonium
entropy at the deconfinement transition can be related to the
entropic force which would induce the dissociation of heavy
quarkonium. In this paper, we study the entropic force in a
rotating hot and dense medium using AdS/CFT correspondence. It
turns out that the inclusion of angular velocity increases the
entropic force thus enhancing quarkonium dissociation, while
chemical potential has the same effect. Furthermore, the results
imply that the quarkonium dissociates easier in rotating medium
compared to static case.

\end{abstract}

\maketitle

\section{Introduction}
In relativistic heavy ion collisions at the relativistic heavy ion
collider (RHIC) and large hadron collider (LHC), a hot and dense,
strongly interacting medium named quark-gluon plasma (QGP) has
been created \cite{JA,KA,EV}. One of the main experimental
signatures of QGP formation is the dissociation of heavy
quarkonium \cite{TMA}. They are expected to be created in the
early stages of the collisions and give us significant information
about the evolving of QGP. Previous research has indicated that
the heavy quarkonium is suppressed due to the Debye screening
induced by the high density of color charges in QGP. But recent
studies of charmonium (heavy quarks of charm and anticharm) show a
puzzle: the charmonium suppression observed at RHIC (lower energy
density) is stronger than at LHC (larger energy density)
\cite{AAD,BBA}. Obviously, this contradicts the Debye screening
scenario \cite{TMA} as well as the thermal activation through the
impact of gluons \cite{DKH,EV1}. Some scholars think that one of
the reasons may be the recombination of the produced charm quarks
into charmonium \cite{PBR,RLT}.

However, recent research suggested \cite{DEK} that the puzzle on
the suppression of the charmonium can be a consequence of the
nature of deconfinement. This argument is based on the lattice QCD
results \cite{DKA1,DKA2,PPE,OK1} indicating a large amount of
entropy $S$ associated with the heavy quark-antiquark pair
($Q\bar{Q}$) around the crossover region of QGP. In particular,
this entropy which grows with the inter-distance $L$ between
$Q\bar{Q}$ leads to the emergent entropic force \cite{DEK}
\begin{equation}
\mathcal{F}=T\frac{\partial S}{\partial L},\label{f}
\end{equation}
where $T$ denotes the temperature of the plasma. It has been shown
that the repulsive entropic force is responsible for dissociating
the quarkonium and then would be a solution of the puzzle.

The AdS/CFT correspondence
\cite{Maldacena:1997re,Gubser:1998bc,EW} or more general
gauge/gravity duality provides a helpful tool to explore various
properties of QGP \cite{JCA,OD0}. With the method,  K. Hashimoto
and D. Kharzeev firstly calculated the entropic force associated
with $Q\bar{Q}$ for $\mathcal N=4$ SYM plasma \cite{KHA}. It was
found that the entropy increases with the inter-quark distance and
the peak of the entropy emerges when the U-shaped string stretched
between the $Q\bar{Q}$ touches the horizon of the black hole.
Subsequently, there have many attempts to address the entropic
force in this direction \cite{KBF,ZQ,II,DE1,ZQ1,ST,MK,MK1,SS,JZ}.

In this paper, we extend the studies of \cite{KHA} to the case of
rotating plasma with chemical potential. In particular, we will
employ the AdS-Reissner Nordstrom (AdS-RN) black hole
\cite{CV:1999,DT:2006} and extend it to a rotating case with
planar horizon. Because there's the strong possibility that the
QGP produced in (typical) noncentral heavy ion collisions may
carry a nonzero angular momentum on the order of
$10^4$-$10^5\hbar$ with local angular velocity in the range of
0.01-0.1 GeV \cite{nat,zt,fb,xg,lg}. Although the major part of
this angular momentum will be taken away by the spectator
nucleons, some amount of angular momentum remains in the QGP
\cite{mib,dek,yj} and thus may give rise to significant observable
effects. On the other hand, the QGP produced in heavy ion
collisions is assumed to carry a finite, albeit small, baryon
number density, e.g., the Beam Energy Scan program at RHIC covers
the beam energies of $\sqrt{S_{NN}}$ = 200, 62.4, 54.4, 39, 27
GeV...corresponding to a region of the chemical potential
$0.025\leq\mu_B\leq0.72 GeV$ \cite{ab}. In this regard, it would
be interesting to study the entropic force at finite temperature
and density under rotation.

This paper is organized as follows: In the next section, we
briefly review the AdS-RN background and extend it to rotating
case. In section 3, we investigate the behavior of the entropic
force in this background and explore how the deconfinement
transition can be viewed as entropic self-destruction. In section
4, we summarize our results and provide a concluding discussion.

\section{background geometry}
From the AdS/CFT correspondence, $\mathcal N=4$ SYM theory with
non-zero chemical potential can be obtained by making the black
hole in the holographic dimension charged. The corresponding
metric is the AdS-RN black hole \cite{CV:1999,DT:2006}
\begin{equation}
ds^2
=-\frac{r^2}{R^2}f(r)dt^2+\frac{r^2}{R^2}d\vec{x}^2+\frac{R^2}{r^2f(r)}dr^2,\label{metric}
\end{equation}
with
\begin{equation}
f(r)=1-(1+Q^2)(\frac{r_h}{r})^4+Q^2(\frac{r_h}{r})^6,
\end{equation}
where $R$ is the curvature radius (for convenience, we set $R=1$
for later discussion), $Q$ denotes the charge of black hole, $r$
refers to the radial coordinate with $r=r_h$ the horizon, defined
by $f(r_h)=0$. The asymptotic boundary is at $r=\infty$. The
string tension $\frac{1}{2\pi\alpha^\prime}$ is related to the 't
Hooft coupling constant $\lambda$ by
$\frac{1}{\alpha^\prime}=\sqrt{\lambda}$.

Following \cite{mb,ce,xc}, one can extend (\ref{metric}) to a
rotating case from the static configuration through a local
Lorentz boost in the $t-\phi$ plane
\begin{equation}
t\rightarrow \gamma(t+\omega l^2 \phi), \qquad \phi\rightarrow
\label{bo} \gamma(\phi+\omega l^2 t), \label{bo}
\end{equation}
with
\begin{equation}
\gamma=\frac{1}{\sqrt{1-\omega^2l^2}},
\end{equation}
where $\phi$ is the angular coordinate describing the rotation.
$\omega$ represents the angular velocity. $l$ is the radius of the
rotating axis. In this work we will focus on the qualitative
results, then we simply set $l=1 GeV^{-1}$, as follows from
\cite{xc}.

Given that, the rotating case of (\ref{metric}) is
\begin{equation}
ds^2=-p(r)dt^2+r^2(dx^2+dy^2)+\frac{1}{r^2f(r)}dr^2+q(r)(d\phi+m(r)dt)^2,\label{metric1}
\end{equation}
with
\begin{equation}
p(r)=\frac{f(r)r^2(1-\omega^2)}{1-f(r)\omega^2},\qquad
q(r)=(1-f(r)\omega^2)r^2\gamma^2, \qquad
m(r)=\frac{\omega(1-f(r))}{1-f(r)\omega^2}.
\end{equation}

The temperature of the black hole reads
\begin{equation}
T=\frac{r_h}{\pi}\sqrt{1-\omega^2}(1-\frac{Q^2}{2}). \label{T}
\end{equation}
where $Q$ is in the range $0\leq Q\leq\sqrt{2}$.

The chemical potential reads \cite{xc}
\begin{equation}
\mu=\sqrt{3}Qr_h\sqrt{1-\omega^2}.\label{mu}
\end{equation}

Notice that the chemical potential implemented here is not the
baryon chemical potential of QCD but a chemical potential
conjugated to the R-charge associated with SYM. However, in such a
context it could serve as a simple way of introducing finite
density effect into the system \cite{ED}.

\section{entropic force in the rotating background}
We now proceed to study the entropic force in the rotating AdS-RN
black hole (\ref{metric1}) following the prescription of
\cite{KHA}. The Nambu-Goto action is
\begin{equation}
S_{NG}=-\frac{1}{2\pi\alpha^\prime}\int d\tau
d\sigma\sqrt{-detg_{\alpha\beta}}, \label{S}
\end{equation}
with
\begin{equation}
g_{\alpha\beta}=g_{\mu\nu}\frac{\partial
X^\mu}{\partial\sigma^\alpha} \frac{\partial
X^\nu}{\partial\sigma^\beta},
\end{equation}
where $g_{\alpha\beta}$ represents the induced metric and
parameterized by $(\tau,\sigma)$ on the string world-sheet,
$g_{\mu\nu}$ is the metric, $X^\mu$ is the target space
coordinate.

Since the transformation (\ref{bo}) is a boost in the $t-\phi$
plane, one may consider the $Q\bar{Q}$ pair located at $x-y$
plane, e.g., one considers the $Q\bar{Q}$ pair to be aligned in
the $x$ direction,
\begin{equation}
t=\tau, \qquad x=\sigma, \qquad y=0,\qquad \phi=0, \qquad
r=r(\sigma). \label{par}
\end{equation}

Based on these assumptions, (\ref{S}) becomes
\begin{equation}
S_{NG}=\frac{\mathcal{T}}{2\pi\alpha^\prime}\int_{-L/2}^{L/2} dx
\sqrt{A(r)+B(r)(\frac{dr}{d\sigma})^2},\label{S0}
\end{equation}
with
\begin{equation}
A(r)=(p(r)-q(r)m^2(r))r^2,\qquad
B(r)=\frac{p(r)-q(r)m^2(r)}{r^2f(r)},
\end{equation}
where the $Q\bar{Q}$ are set at $x=-L/2$ and $x=L/2$,
respectively.

Since (\ref{S0}) does not depend on $\sigma$ explicitly, one
obtains a conserved quantity,
\begin{equation}
\mathcal L-\frac{\partial\mathcal
L}{\partial\dot{r}}\dot{r}=constant.
\end{equation}

Imposing the boundary condition (the deepest point of the U-shaped
string)
\begin{equation}
\dot{r}=0,\qquad  r=r_c \qquad (r_h<r_c)\label{con},
\end{equation}
one gets
\begin{equation}
\frac{dr}{d\sigma}=\sqrt{\frac{A^2(r)-A(r)A(r_c)}{A(r_c)B(r)}}\label{dotr},
\end{equation}
with $A(r_c)=A(r)|_{r=r_c}$.

Integrating (\ref{dotr}), the inter-distance of $Q\bar{Q}$ is
found to be
\begin{equation}
L=2\int_{r_c}^{\infty}dr\sqrt{\frac{A(r_c)B(r)}{A^2(r)-A(r)A(r_c)}}\label{x}.
\end{equation}

On the other hand, the entropy is given by
\begin{equation}
S=-\frac{\partial F}{\partial T},\label{s}
\end{equation}
where $F$ denotes the free energy of $Q\bar{Q}$. Notice that $F$
has been studied at zero temperature \cite{JMM} and finite
temperature \cite{ABR,SJR} from holography. Generally, there are
two situations:

1. If $L>\frac{c}{T}$ (where $c$ denotes the maximum value of
$LT$), some new configurations need to be taken into account and
thus there are several alternatives for $F$ \cite{DB,MCH}. If one
selects a configuration of two disconnected trailing drag strings
\cite{CPH,GB}, the corresponding free energy can be written as
\begin{equation}
F^{(1)}=\frac{1}{\pi\alpha^\prime}\int_{r_h}^{\infty}dr,
\end{equation}
leads to
\begin{equation}
\mathcal{S}^{(1)}=\sqrt{\lambda}\theta(L-\frac{c}{T})\label{S1},
\end{equation}
where $\theta(L-\frac{c}{T})$ is the Heaviside step function.

2. If $x<\frac{c}{T}$, the fundamental string is connected. Then
$F$ can be obtained from the on-shell action of the fundamental
string in the dual geometry,
\begin{equation}
F^{(2)}=\frac{1}{\pi\alpha^\prime}\int_{r_c}^{\infty} dr
\sqrt{\frac{A(r)B(r)}{A(r)-A(r_c)}}.\label{S2}
\end{equation}

Then from (\ref{T}), (\ref{s}) and (\ref{S2}), one gets
\begin{equation}
S^{(2)}=-\frac{\partial F^{(2)}}{\partial
T}=-\frac{1}{2\alpha^\prime}\frac{1}{\sqrt{1-\omega^2}(1-\frac{Q^2}{2})}\int_{r_c}^{\infty}
dr\frac{[A^\prime(r)B(r)+A(r)B^\prime(r)][A(r)-A(r_c)]-A(r)B(r)[A^\prime(r)-A^\prime(r_c)]}{\sqrt{A(r)B(r)[A(r)-A(r_c)]^3}}\label{S21},\label{fo}
\end{equation}
with
\begin{eqnarray}
A^\prime(r)&=&r^2(p^\prime-q^\prime m^2-2qmm^\prime), \nonumber\\
B^\prime(r)&=&\frac{(p^\prime-q^\prime m^2-2qmm^\prime)f-(p-qm^2)f^\prime}{r^2f^2},\nonumber\\
p^\prime&=&\frac{r^2(1-\omega^2)(f^\prime(1-f\omega^2)+\omega^2ff^\prime)}{(1-f\omega^2)^2},\nonumber\\
q^\prime&=&-r^2\gamma^2\omega^2f^\prime,\nonumber\\
m^\prime&=&\frac{\omega(-f^\prime(1-f\omega^2)+\omega^2(1-f)f^\prime)}{(1-f\omega^2)^2},\nonumber\\
f^\prime&=&-4(1+Q^2)r_h^3r^{-4}+6Q^2r_h^5r^{-6},
\end{eqnarray}
where $A^\prime(r_c)\equiv A^\prime(r)|_{r=r_c}, q\equiv q(r),
m\equiv m(r), f\equiv f(r)$, and the derivatives are with respect
to $r_h$. One can check that by setting $\omega=0$ and $\mu=0$ (or
$Q=0$) in (\ref{fo}), the result of SYM \cite{KHA} will be
recovered.

\begin{figure}
\centering
\includegraphics[width=8cm]{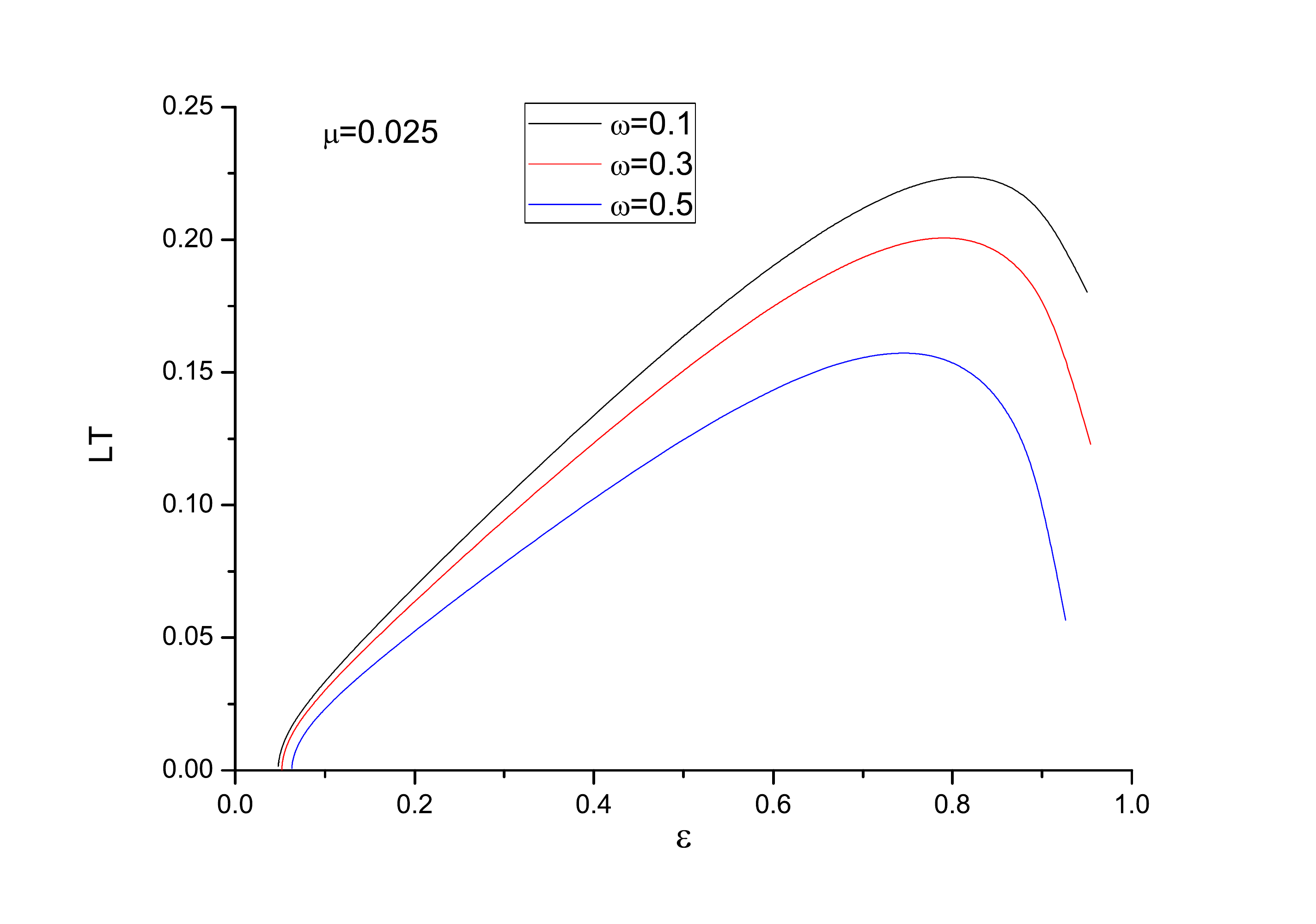}
\includegraphics[width=8cm]{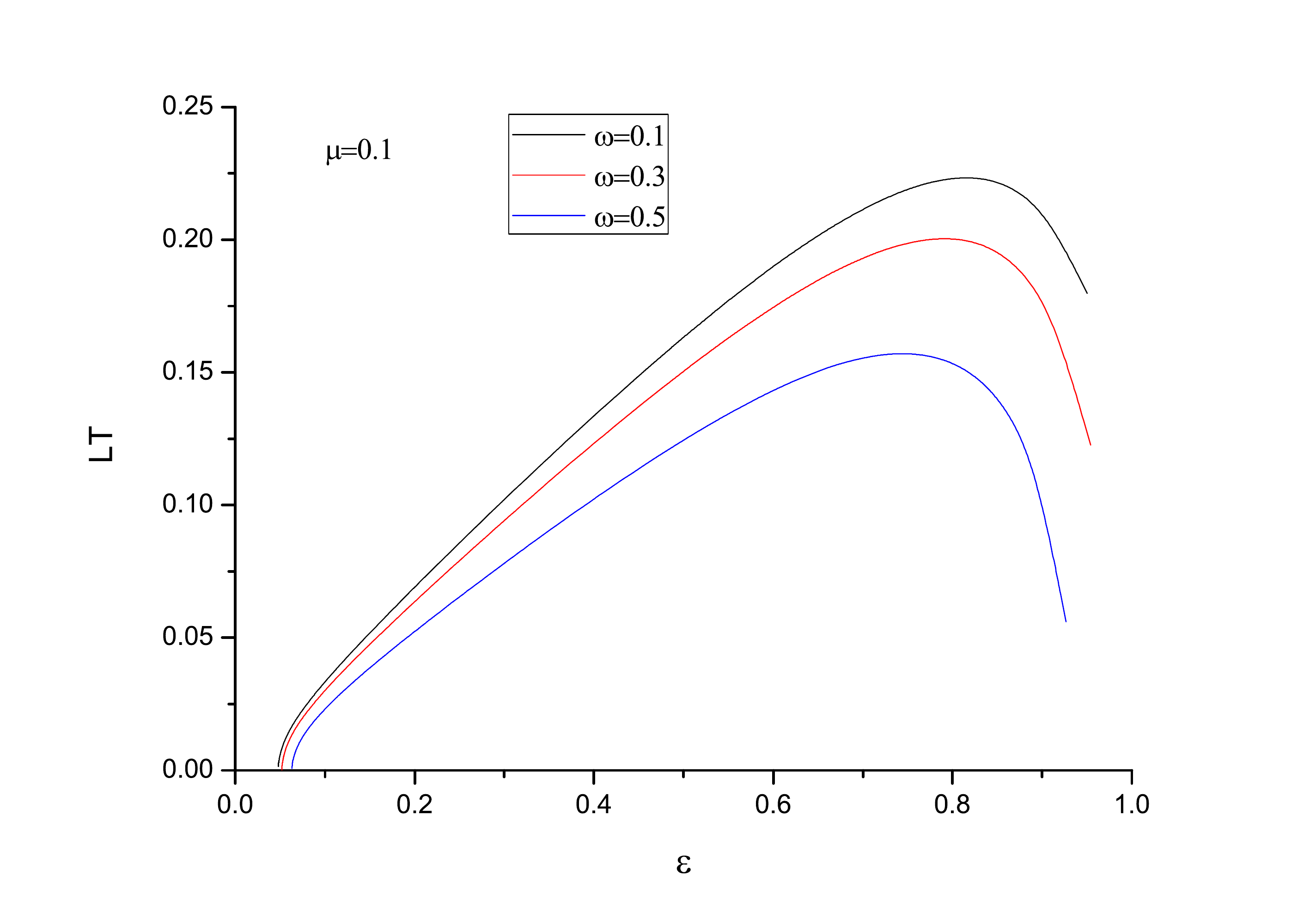}
\caption{$LT$ versus $\varepsilon$ for different $\omega$. Left:
$\mu=0.025GeV$; Right: $\mu=0.1GeV$. In both panels from top to
bottom $\omega=0,0.1,0.3GeV$, respectively.}
\end{figure}

\begin{figure}
\centering
\includegraphics[width=8cm]{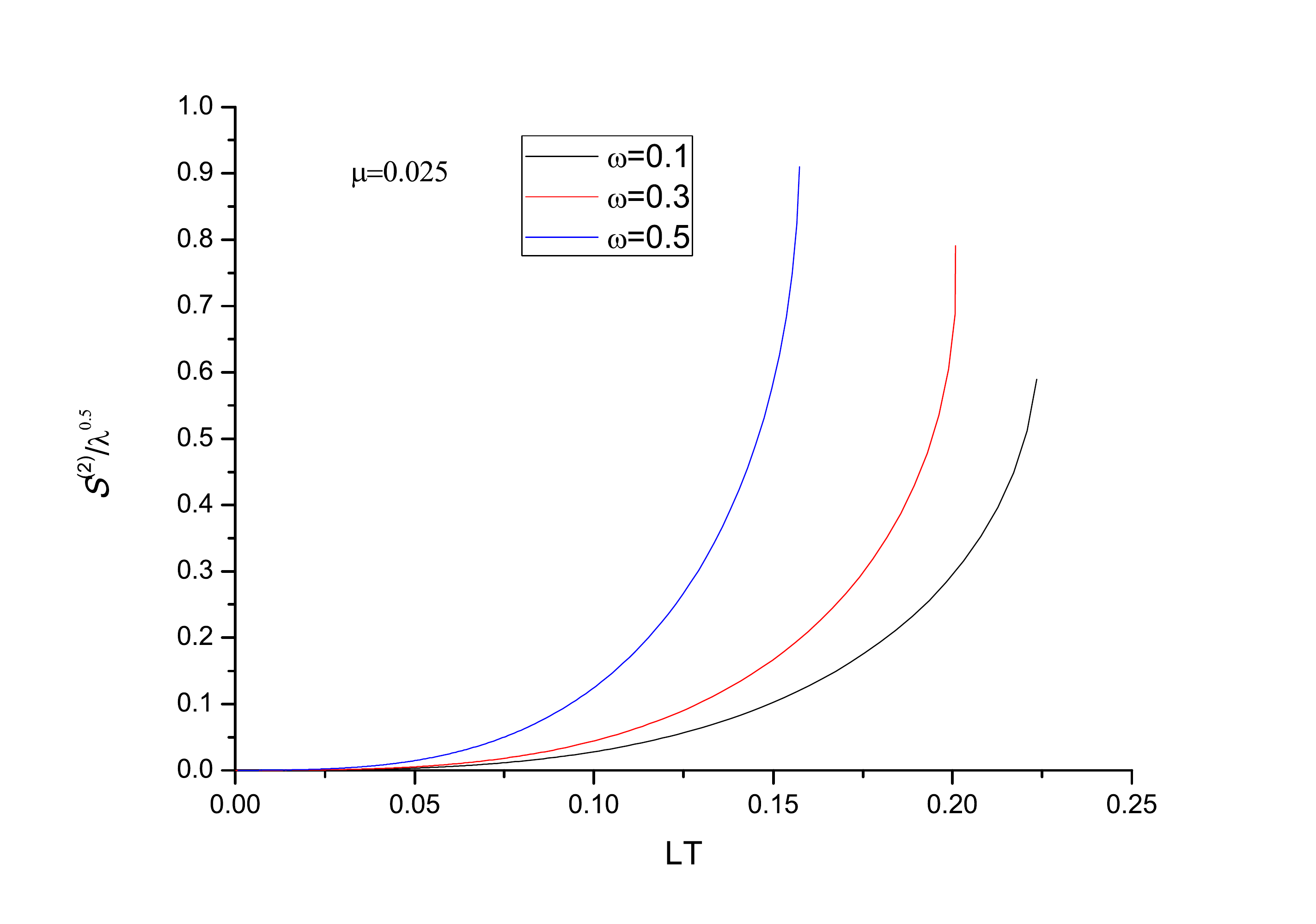}
\includegraphics[width=8cm]{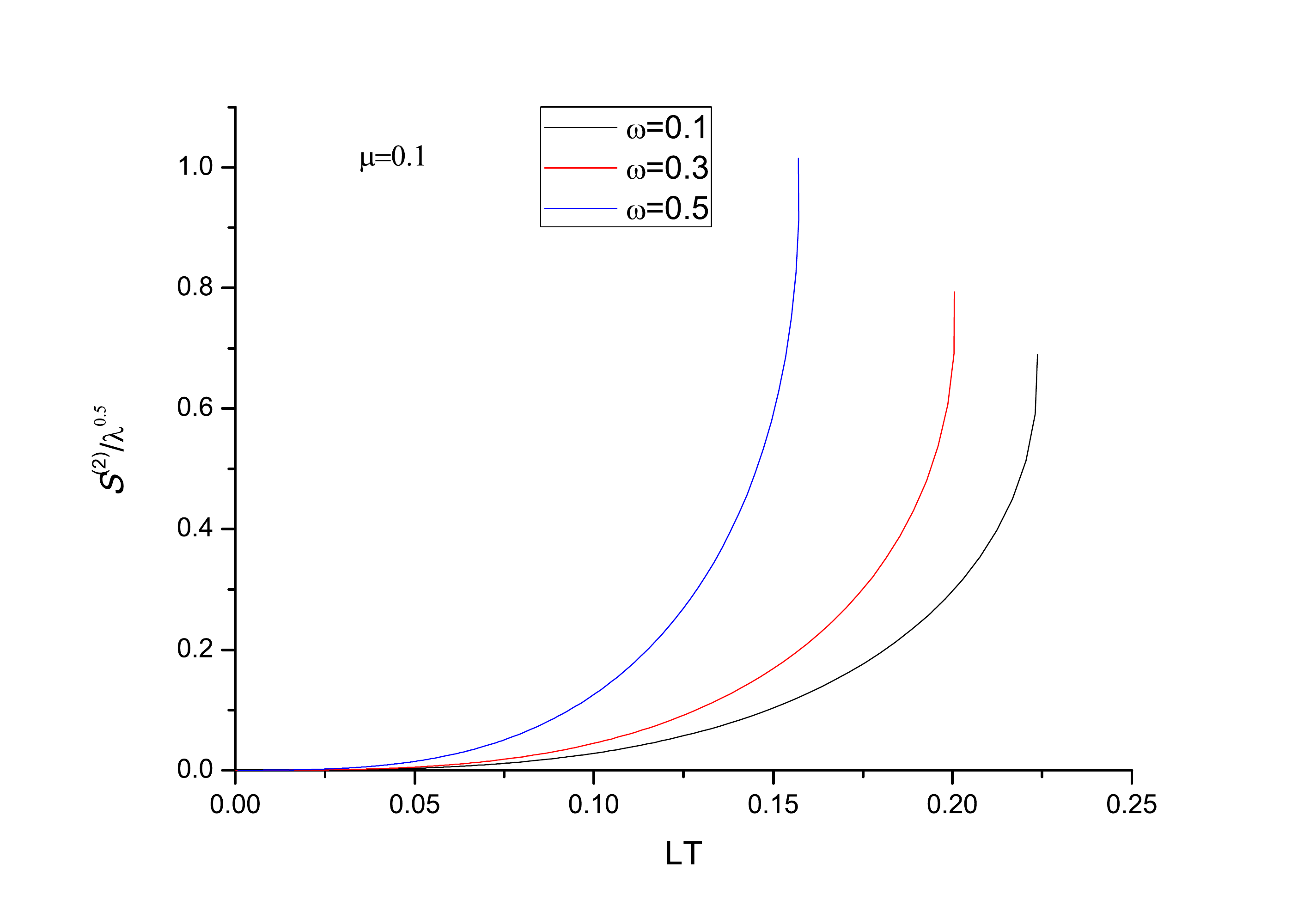}
\caption{$S^{(2)}/\sqrt{\lambda}$ versus $LT$ for different
$\omega$. Left: $\mu=0.025GeV$; Right: $\mu=0.1GeV$. In both
panels from right to left, $\omega=0,0.1,0.3GeV$, respectively.}
\end{figure}

\begin{figure}
\centering
\includegraphics[width=8cm]{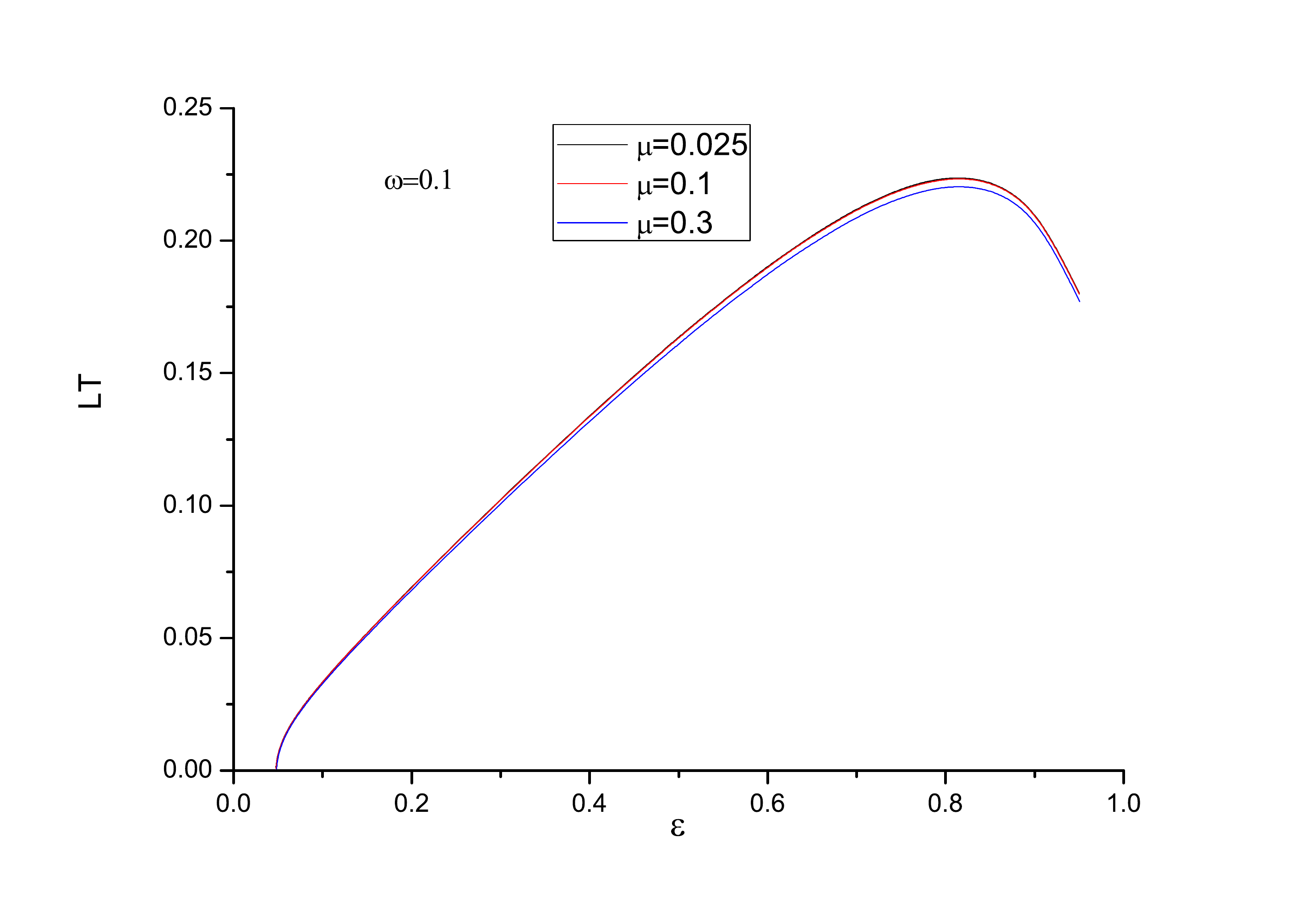}
\includegraphics[width=8cm]{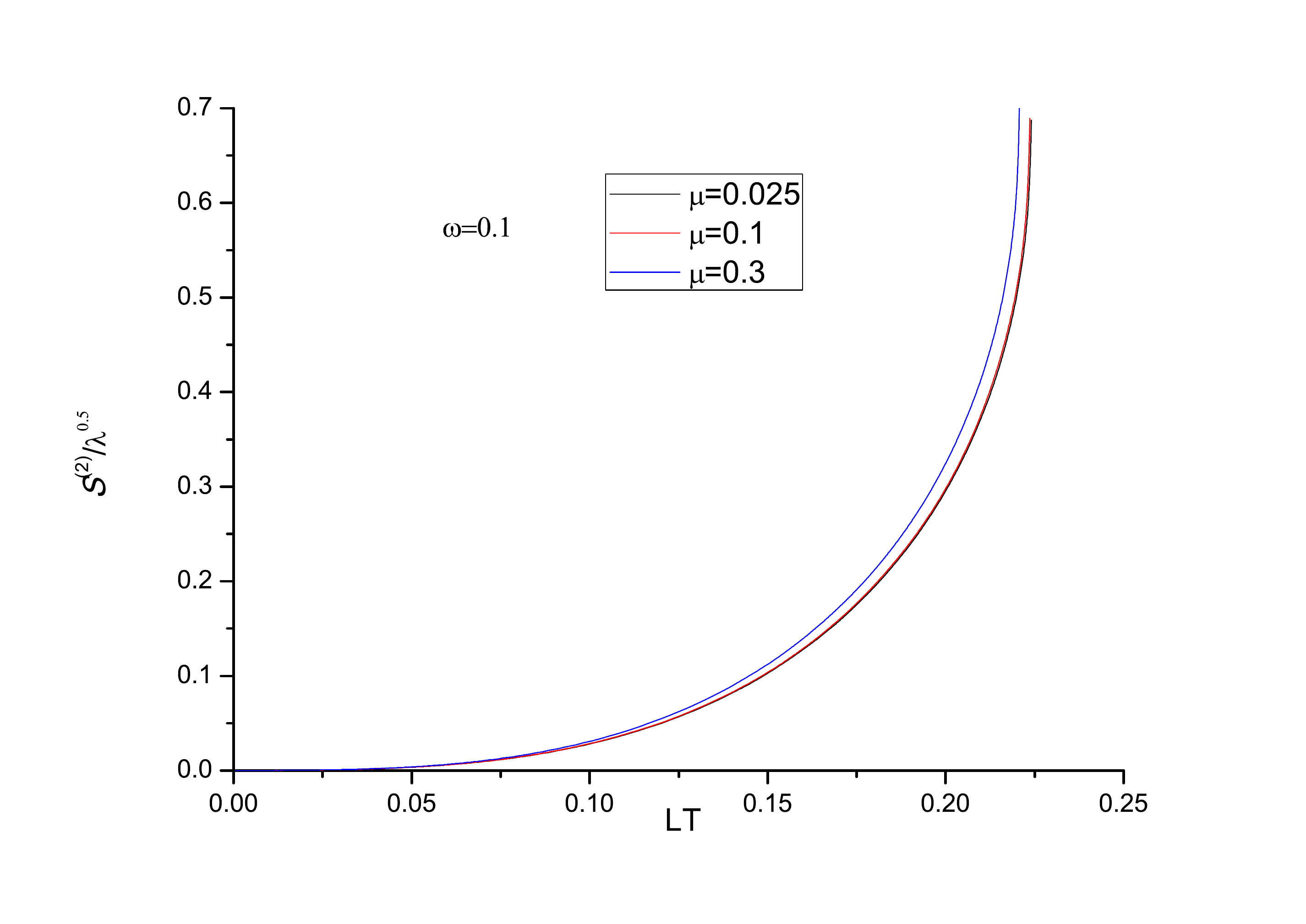}
\caption{Left: $LT$ versus $\varepsilon$ for different $\mu$, from
top to bottom $\mu=0.025,0.1,0.3GeV$, respectively. Right:
$S^{(2)}/\sqrt{\lambda}$ versus $LT$ for different $\mu$, from
right to left, $\mu=0.025,0.1,0.3GeV$, respectively. In both cases
we take $\omega=0.1 GeV$. }
\end{figure}

Let's discuss results. First, we analyze how angular velocity
affects the inter-distance of $Q\bar{Q}$. To this end, we plot
$LT$ as a function of $\varepsilon\equiv r_h/r_c$ for various
values of $\omega$ in fig.1, where the left panel is for
$\mu=0.025 GeV$ while the right $\mu=0.1 GeV$ (notice that in all
the plots $\omega$ and $\mu$ are in unit GeV which are not
mentioned for the brevity of the notation)). In both panels from
top to bottom $\omega=0.1,0.3,0.5GeV$, respectively. From each
panel, one can see that by increasing $\omega$, $LT$ decreases.
Namely, the inclusion of angular velocity reduces the
inter-distance.

Moreover, to see how angular velocity modifies the entropic force,
we plot $S^{(2)}/\sqrt{\lambda}$ versus $LT$ for various cases in
fig.2. Similarly, one chooses $\mu=0.025, 0.1 GeV$ and
$\omega=0.1,0.3,0.5GeV$ in calculations. From these figures, one
sees that increasing $\omega$ leads to larger entropy at small
distances. It is known that the entropic force (see Eq.(\ref{f}))
depends on the growth of the entropy with the inter-distance and
is responsible for dissociating the quarkonium. Therefore, one can
draw the conclusion that the inclusion of angular velocity
increases the entropic force thus enhancing the quarkonium
dissociation, in agreement with \cite{JZ}.

Also, one can analyze the chemical potential dependence of the
entropic force. For this purpose, we plot $LT$ versus
$\varepsilon$ and $S^{(2)}/\sqrt{\lambda}$ versus $LT$ for
different values of $\mu$ in fig.3. From these figures, one finds
the presences of chemical potential also reduces the
inter-distance and enhances the entropic force, consistent with
\cite{ZQ}. The physical significance of the results will be
discussed in the final section.

\section{conclusion}
Recent studies have shown \cite{KHA} that the peak of the
quarkonium entropy at the deconfinement transition can be related
to the entropic force which can destruct the heavy quarkonium. In
this paper, we extended the studies of \cite{KHA} to the case of
rotating medium with chemical potential using AdS/CFT
correspondence. It is shown that the inclusion of angular velocity
increases the entropic force thus enhancing the quarkonium
dissociation, while chemical potential has the same effect.
Furthermore, the results imply that quarkonium dissociates easier
in rotating medium compared to static case.

Interestingly, the entropic force of a moving heavy quarkonium has
been studied in \cite{KBF} and the results show that the entropic
force destructs the moving quarkonium easier than the static case.
Since motion is relative, their results can be understood as:
quarkonium dissociates easier in moving medium compared to static
case. If compare their results with ours, one may infer that
translation and rotation have the same effect on quarkonium
dissociation. In particular, quarkonium dissociates easier in
moving or rotating medium compared to static case.

However, there are some inadequacies in this research, e.g, the
model we employed here is not a consistent model. Considering the
entropic force in some consistent models, e.g.
\cite{JP,AST,DL1,DL,SH,SH1} would be instructive. On the other
hand, the entropic force mechanism applies only to charmonium, but
hardly applies to bottomonium (the mass of $c$ quark is about 1.27
GeV, while $b$ quark 4.2 GeV). It was argued \cite{DEK} that most
of the bottomonium states have smaller sizes, which are much less
influenced by the entropic force.

Finally, it's worth noting that the rotating QGP can also be
described by means of five-dimensional Kerr-AdS black hole
\cite{sw}. Then one can study the entropic force in that rotating
frame as well. It will be left as a further study.

\section{Acknowledgments}
This work is supported by the NSFC under Grant Nos. 11805052,
12147219.

\end{document}